\documentclass{emulateapj}

\begin{document}

\title{Long-term variations in the pulse emission from PSR
J0737$-$3039B} 

\author{M. Burgay\altaffilmark{1}, A. Possenti\altaffilmark{1},
R. N. Manchester\altaffilmark{2}, M. Kramer\altaffilmark{3},
M. A. McLaughlin\altaffilmark{3}, D. R. Lorimer\altaffilmark{3},
I. H. Stairs\altaffilmark{4}, B. C. Joshi\altaffilmark{5},
A. G. Lyne\altaffilmark{3}, F.  Camilo\altaffilmark{6},
N. D'Amico\altaffilmark{7}, P. C. C. Freire\altaffilmark{8},
J. M. Sarkissian\altaffilmark{2}, A. W. Hotan\altaffilmark{9}, and
G. B. Hobbs\altaffilmark{2}}

\altaffiltext{1}{INAF-Osservatorio Astronomico di Cagliari, Loc.
Poggio dei Pini, Strada 54, 09012 Capoterra, Italy}
\altaffiltext{2}{Australia Telescope National Facility, CSIRO,
P.O. Box 76, Epping, NSW 1710, Australia.}
\altaffiltext{3}{Jodrell Bank Observatory, University of Manchester,
Macclesfield, Cheshire SK11 9DL, UK.}
\altaffiltext{4}{Department of Physics and Astronomy, University of British
Columbia, 6224 Agricultural Road, Vancouver, BC V6T 1Z1, Canada.}
\altaffiltext{5}{National Centre for Radio Astrophysics, P.O. Bag 3, 
Ganeshkhind, Pune 411 007, India.}
\altaffiltext{6}{Columbia Astrophysics Laboratory, Columbia University,
550 West 120th Street, New York, NY 10027.}
\altaffiltext{7}{Universit\`a degli Studi di
Cagliari, Dipartimento di Fisica, SP Monserrato-Sestu km
0.7, 09042 Monserrato, Italy.}
\altaffiltext{8}{National Astronomy and Ionosphere Center, Arecibo 
Observatory, HC-3, Box 53995, PR 00612.}
\altaffiltext{9}{Swinburne Centre for Astrophysics and Supercomputing,
Mail 31, P.O. Box 218, Hawthorn, VIC 3122, Australia.}

\shorttitle{Pulse Emission from PSR J0737 3039B}
\shortauthors{Burgay et al.}
\slugcomment{Received 2005 February 24; accepted 2005 April 1; published
  2005 April 15}
\journalinfo{The Astrophysical Journal, 624:L113 L116, 2005 May 10}

\begin{abstract}
Analysis of 20 months of observations at the Parkes radio telescope
shows secular changes in the pulsed emission from J0737$-$3039B, the
2.77 s pulsar of the double-pulsar system. Pulse profiles are becoming
single-peaked in both bright phases of the orbital modulation although
there is no clear variation in overall pulse width.  The shape of the
orbital modulation is also varying systematically, with both bright
phases shrinking in longitude by $\sim 7^{\circ}$ yr$^{-1}$. However,
the combined span of the two bright phases is relatively constant and
together they are shifting to higher longitudes at a rate of $\sim
3^{\circ}$ yr$^{-1}.$ We discuss the possible contributions of
geodetic precession and periastron advance to the observed variations.
\end{abstract}


\section{Introduction} \label{sec:intro}

One of the many unique features of the double-pulsar system
J0737$-$3039A/B \citep{bdp+03,lbk+04} is the dramatic orbital modulation
of the pulsed emission from PSR J0737$-$3039B (hereafter ``B''), the
2.77 s pulsar of the system. Both the shape and the intensity of the B
pulse profile vary with orbital longitude, with two bright phases
centered around longitudes (with respect to the ascending node) of
$\sim 210\degr$ (bright phase 1, hereafter bp1) and $\sim 280\degr$
(hereafter bp2), respectively. The pulse is weakly visible at other orbital
longitudes, except perhaps between $30\degr$ and $60\degr$.

This unprecedented behavior has been interpreted as being due to the
interaction between radiation from the 23 ms pulsar J0737$-$3039A (hereafter
``A'') and B's magnetosphere \citep{jr04,zl04,lut04,lut05}. Direct
evidence of the mutual interaction between the two pulsars is
manifested by the drifting behavior at A's period in the single pulses
of B \citep{mkl+04} and by the modulation at B's period of the
emission from A during its eclipse \citep{mll+04}.

The geometry of the system is expected to vary with time as a result
of the strong relativistic effects occurring in the system: the geodetic
precession \citep{dr74} and periastron advance of B have periods of
only 71 and 21 yr, respectively, both much less than for any other
known pulsar binary system \citep{lbk+04}. These effects are expected
to result in secular changes in the observed pulse shape
\citep[see e.g.][]{kra98}, in the interaction between the two pulsars, and
hence in the orbital modulation. Somewhat surprisingly, Parkes
observations of A show no significant evidence of profile-shape
variation over a 3 yr interval \citep{mkp+05}. Near alignment of
A's rotation axis with the orbit normal is a possible explanation for
the observed lack of variations.

In this Letter we report the detection of secular variations in both
B's pulse shape and the orbital modulation of B's emission using
observations with the Parkes radio telescope spanning 20 months. In \S
\ref{sec:obs} we present the observations and results and in \S
\ref{sec:discussion} we discuss their interpretation.

\section{Observations and Results}
\label{sec:obs}

The double-pulsar system J0737$-$3039A/B was observed at 20~cm (1390 MHz)
using the Parkes 64 m radio telescope from 2003 May 1 to 2005 January
1. All observations were made using the central beam of the multibeam
receiver, except from November 2003 to September 2004 when the ``H-OH''
receiver was used. A filter-bank system providing 512 total-power
channels across 256 MHz was used. Further details of the observing
systems are given in \citet{mkp+05}. For the present analysis we used
a total of 94 hr of data spread over 105 observations, more or less
uniformly distributed across the 20 month data span.

\begin{figure*}[ht]
\centerline{\includegraphics[angle=270,width=150mm]{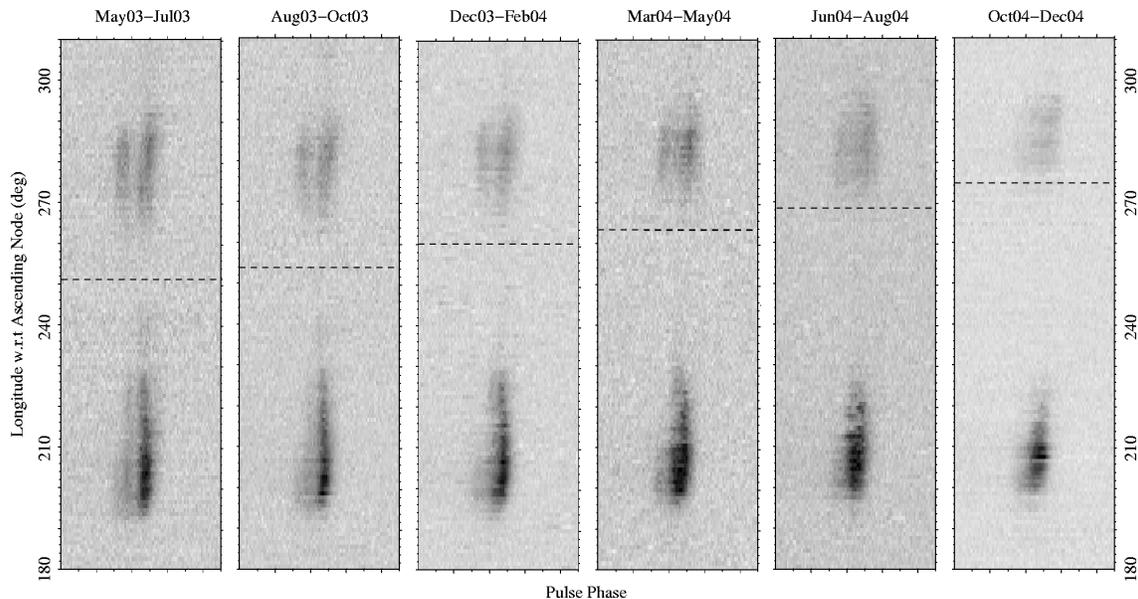}}
\caption{Intensity of the PSR J0737$-$3039B pulse emission at 1390 MHz
as a function of orbital longitude and pulse phase at six epochs. Only
the longitude range $180\degr$-$310\degr$ covering the two bright
phases bp1 ({\it{lower}}) and bp2 ({\it{upper}}) and a pulse-phase
window of 0.1 pulse periods centered on the pulse are shown. The
dashed lines represent the longitude of periastron at each epoch.}
\label{fig:fig1}
\end{figure*} 

Figure~\ref{fig:fig1} shows the intensity of the two bright phases of
emission from B as a function of the orbital longitude with respect to
the ascending node for six epochs across the 20 month data span. Each
panel was obtained by summing the observations taken over an
$\sim$3 month interval. The PSRCHIVE data analysis system
\citep{hvm04} was used to sum and image the data. These plots confirm
the overall picture presented in \citet{lbk+04} and \citet{rbd+05},
showing the complex pulse shape and flux density variations as a
function of orbital longitude. Furthermore, Figure \ref{fig:fig1}
clearly shows that both the extent in orbital longitude of the bright
phases and the shape of the pulse at a given orbital longitude are
evolving with time. We also note that the mean pulse intensity in bp2 is
decreasing relative to that in bp1. In the following sections we
quantify these trends.

\subsection{Pulse Profile Variations}
\begin{figure}[ht]
\centerline{\includegraphics[scale=0.4]{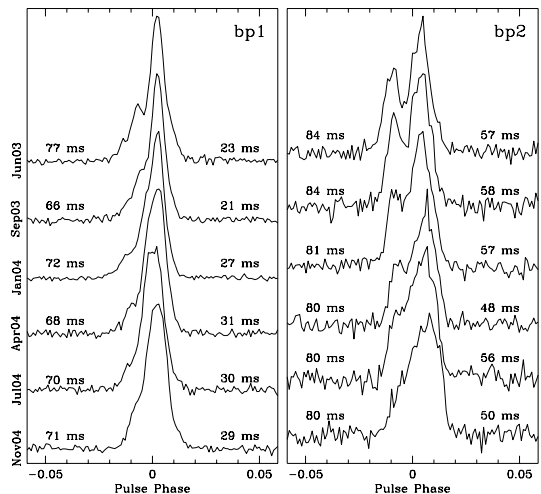}}
\caption{Mean pulse profiles, normalized in peak, of PSR J0737$-$3039B
at 1390 MHz for the two bright phases, bp1 ({\it{left}}, orbital longitude
$190\degr$-$235\degr$) and bp2 ({\it{right}}, orbital longitude 
$260\degr$-$300\degr$) at six epochs from 2003 June ({\it{top}}) to
2004 November ({\it{bottom}}). All profiles have been scaled to the same peak
height. Numbers near each profile are the pulse widths at 10\% ({\it{left}})
and 50\% ({\it{right}}) of the pulse peak. For bp1, these widths have an
uncertainty of about 1 ms, and for bp2, about 3 ms. }\label{fig:fig2}
\end{figure}

As Figure~\ref{fig:fig1} shows, the mean pulse profile for B is quite
different in bp1 and bp2 and even varies somewhat with orbital
longitude within each bright phase. Despite the small variations
within each bright phase, we have chosen to illustrate the trends in
profile shape by integrating across bp1 and bp2 separately for each of
the epochs corresponding to those in Figure~\ref{fig:fig1}. The
resulting pulse profiles are shown in Figure \ref{fig:fig2}. This
figure shows significant long-term variations in the mean pulse
profile for both bright phases. The variations are in fact very
similar in both bp1 and bp2, with the leading component getting weaker
and the trailing component becoming broader with time. There is little
or no significant evolution in the 10\% width of either profile (upper
limits of 7 and 8 ms yr$^{-1}$ for bp1 and bp2, respectively). For bp1,
the width of the trailing component appears to increase with time, but
the change is nonlinear, with the first two profiles narrow and the
last four broader. The feature on the leading edge of the trailing
component in the 2004 January profile suggests that this width
increase is due to the relatively rapid growth of an additional
component between the two other components. For bp2 the 50\% widths
are difficult to interpret as the leading component has a relative
amplitude close to 50\% and the signal-to-noise ratio is smaller.

\subsection{Orbital Phase Variations}

\begin{figure}[ht]
\centerline{\includegraphics[scale=0.4]{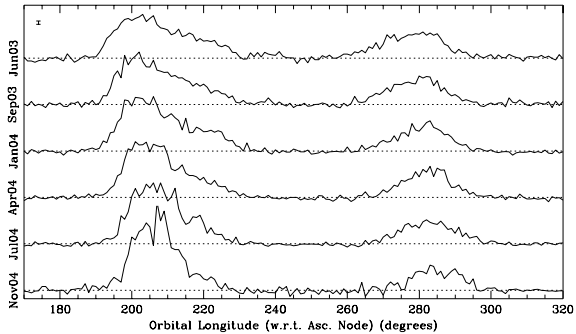}}
\caption{Variation of pulse intensity as a function of orbital
longitude for six epochs. The small bar represents the typical rms noise in
the baseline. }\label{fig:orbvar}
\end{figure}

To quantify the orbital evolution of the bright phases, we took the
data illustrated in Figure~\ref{fig:fig1} and summed the pulse
intensity at each epoch over the 3\% of pulse phase where the pulse is
present. A baseline level calculated in the orbital longitude range
$30\degr$ -- $60\degr$ was then subtracted, and the area under the
resulting curves equalized to remove the effects of refractive
interstellar scintillation. The resulting normalized pulse intensity
curves are shown in Figure~\ref{fig:orbvar} for each of the six
epochs.

\begin{figure}[ht]
\centerline{\includegraphics[scale=0.35,angle=0]{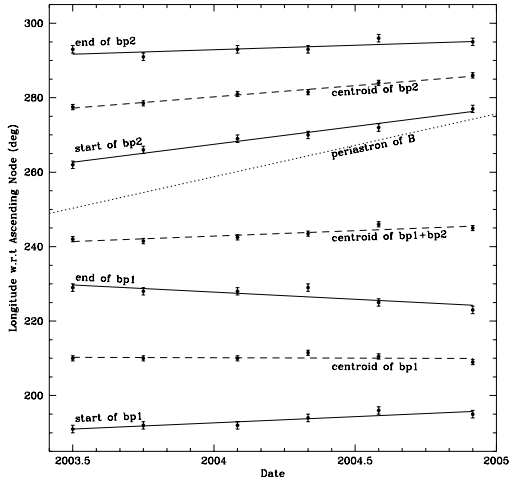}}
\caption{Variations with time of the start and end of the two bright
  phases, measured at the 10\% points, and the centroids of the
  two phases separately and together. The dotted line shows the
  precession of periastron during the 20-month interval. }\label{fig:orbphs}
\end{figure}

\begin{figure}[ht]
\centerline{\includegraphics[scale=0.35,angle=0]{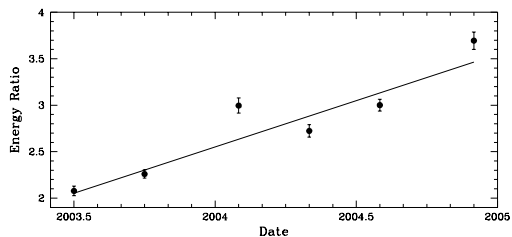}}
\caption{Variation with time of flux(bp1/bp2), the relative integrated
   pulse intensity of the two bright phases.}\label{fig:bratio}
\end{figure}

This figure shows clear systematic changes in the orbital modulation
with time. The most obvious change is that the saddle region between
the two bright phases is becoming less filled: that is, the end of bp1
and the start of bp2 are becoming more separated. Less obvious are that
the modulation as a whole is moving toward later longitudes and that
relative to bp1, bp2 is becoming fainter. These changes are shown more
quantitatively in Figures~\ref{fig:orbphs} and \ref{fig:bratio}. 


Linear least-squares fits to the points in Figure~\ref{fig:orbphs}
show that the starting point of bp1 and the end of bp2 are slowly
shifting toward higher longitudes at a rate of $3\fdg4 \pm 1\fdg7$
and $2\fdg5 \pm 1\fdg7$ yr$^{-1}$, respectively (all the
errors are quoted at 2 $\sigma$). Likewise, the centroid of bp1 and bp2
taken together is moving at a rate of $3\fdg0 \pm 1\fdg2$
yr$^{-1}$, confirming that the overall modulation pattern is shifting
toward later longitudes. However, the interior points have a different
behavior, with the end of bp1 moving backward at $-3\fdg8 \pm
1\fdg7$ yr$^{-1}$ and the start of bp2 advancing at $9\fdg6 \pm
1\fdg7$ yr$^{-1}$. This shows that bp1 and bp2 are shrinking at
a rate of $7\fdg2 \pm 2\fdg4$ and $7\fdg1 \pm 2\fdg4$
yr$^{-1}$, respectively, and the gap between the two bright phases is
widening at $13\fdg4 \pm 2\fdg4$ yr$^{-1}$.

Figure~\ref{fig:bratio} shows the relative energy in bp1 and bp2,
obtained by integrating under the curves in Figure~\ref{fig:orbvar},
from $190\degr$ to $235\degr$ for bp1 and from $260\degr$ to $300\degr$ for
bp2, and taking the ratio. Clearly, bp2 is getting fainter relative to
bp1; a linear fit to the ratio has a slope of $1.0 \pm 0.1$ yr$^{-1}$. 

\section{Discussion}
\label{sec:discussion}

This Letter reports the first observation of secular changes in the pulsed
emission from the J0737$-$3039A/B binary system. The B pulse profile,
initially consisting of two obvious components in both bright phases,
has become single peaked over the 20-month interval of our
observations.  This constrasts with the apparent stability of A's
profile over a similar interval \citep{mkp+05}. Moreover, the orbital
extent of both bright phases is shrinking at a rate of $\sim 7\degr$
yr$^{-1}$ and bp2 is getting progressively fainter with respect to
bp1. Both the outer edges and the centroid of the two bright phases
taken together are advancing in orbital longitude at a rate of $\sim
3^{\circ}$ yr$^{-1}$, whereas there is no significant variation in the
separation of the outer edges of the modulation pattern. All these
numbers have been derived by analyzing observations made at 1.4 GHz;
similar trends appear in a preliminary inspection of unpublished data
taken at the Green Bank and the Giant Metrewave Radio Telescopes at
lower frequencies.
 
If these changes were to continue at the current rate, the second bright
phase would disappear by the end of 2007 when its centroid will be at
orbital longitude $\sim 305^{\circ}$, whereas bp1 will fade away in
mid-2009, with the centroid at orbital longitude $\sim 210^{\circ}$.
We note that the shape of the integrated pulse profile of B at orbital
longitudes away from the two bright phases is single-peaked with 10\%
and 50\% widths ($\sim 65$ and $\sim 30$ ms, respectively) similar to
the corresponding values for the pulse profile in bp1. This may indicate
that B's emission is tending to become more uniform along the orbit as 
the physical processes producing the bright regions cease to operate. 

Provided B's spin and total angular momentum vectors are not aligned,
general relativity predicts \citep[e.g.,][]{bo75} that B's rotational
axis will precess about the orbit normal at a rate $\Omega_g=5\fdg1$
yr$^{-1}$. This precession will vary the angle of impact of A's wind
on B's open magnetic field lines. Several authors
\citep{lbk+04,jr04,lut04,zl04,lut05} have suggested that the pulsed
emission from B is modulated by the impinging energetic flux from A,
so some kind of evolution in B's orbital light curve may be
expected. We note that $\Omega_g$ is of the same order of magnitude as
the rate of the observed variations in the location and extent of bp1
and bp2.

Geodetic precession will also change our line of sight across B's
emission cone, affecting the shape and width of the observed pulse
profile as reported for other pulsars \citep{wrt89,kra98,sta04}. In
fact, the profile shapes in both bp1 and bp2 are clearly
evolving. However, there is no significant variation in the angular
extent of the pulse emission, with limits of about $1\degr$ yr$^{-1}$
on the change of the 10\% widths. This may indicate that B's spin axis
is almost parallel to orbit normal, possibly supporting the hypothesis
that wind-torques from A, which dominate the energetics of the system,
have caused the spin axis of B to almost align with the direction of
the orbital angular momentum \citep{absk05,drb+04}.
 
Periastron precession has moved the orbit with respect to the
ascending node by $28^{\circ}$ over our data span. This large angular
shift is not reflected in any of the measured variations of the
orbital light curves and so the relative position of the two stars
with respect to periastron cannot be a significant factor in the
changing orbital modulation of the emission from B. Likewise, it
cannot be responsible for the shrinking width of the bright phases. In
fact, despite the longitudinal extents of bp1 and bp2 undergoing a
similar decreasing trend, the linear distance between the two pulsars has
changed with opposite signs in bp1 and bp2 along the data span: an
$\sim 3\%$ increment for bp1 and an $\sim 1\%$ decrement for bp2.
 
Various models have been proposed to explain the dramatic orbital
modulation of the B pulse emission. \citet{jr04} suggested that B
brightens only when the emission beam from A illuminates B's
magnetosphere. In its present form, the model is inconsistent with the
observed stability of A's emission \citep{mkp+05}, and it does not
include B's geodetic precession or the effect of the varying
separation of the two pulsars at a given orbital
longitude. 

\citet{zl04} propose that the modulation is a consequence of particles
from A's wind streaming into some of the open magnetic field lines of
B. The model requires that A's wind is anisotropic and also that B's
spin axis is not parallel to the orbital angular momentum vector, thus
allowing geodetic precession to have an effect: the size and the
location of the portion of B's open magnetosphere directly seen by A's
wind at any given orbital longitude would be modified.

Alternatively, \citet{lut04} first suggested that B brightens when the
line of sight runs parallel to the magnetic field lines in the portion
of the B's magnetosphere where the field lines are swept back owing to
the impinging A's wind: in fact, strongly curved magnetic field lines
would favor the growth of a Cerenkov-drift instability, producing
radio emission beamed along (and polarized perpendicular to the plane
of) the magnetic field lines. In this model geodetic precession
affects the location along the orbit (and the spin phase of B) at
which the condition of alignment between the bent-back field lines and
the line of sight is met.

More recently, \citet{lut05} has proposed that the modulations of the
B light curve are due to orbital phase-dependent distortions of B's
magnetosphere by A's wind. In this model, B is intrinsically bright at
all orbital phases, but strong emission is detected only at those
orbital longitudes where the magnetic field lines along which the
radio emission is generated are bent toward the observer.

Assuming the geometry inferred in \citet[$\alpha \sim 75^\circ$
and $\delta \sim 60^\circ$, where $\alpha$ is the angle between spin
and magnetic axes of B, and $\delta$ is the angle between spin axis and
orbit normal]{lt05}, this model roughly accounts for both the extent and the
location of the two bright phases. It can also account for the
variations of the pulse shape along the orbit \citep{lbk+04,rbd+05}
and predicts that the separation between bp1 and bp2 will vary
because of geodetic precession. The magnitude and direction of the
changes we observe will place further constraints on this model
parameters.
 
With continued monitoring of the secular evolution of the emission
from the J0737$-$3039A/B double-pulsar system, including the
measurement of the changes in polarization properties, we should be
able to remove the existing degeneracy both in the orientation of the
spin and magnetic axes of the two pulsars and among various
precessional effects; this will also be crucial for measuring the moment
of inertia of pulsar A once we measure higher-order terms to
periastron advance at the second Post-Newtonian level
\citep{klb+05}. Combining these data with an improved understanding of
the complicated magnetospheric interactions and emission physics in
the system, we may hope to eventually obtain an explanation for this
most intriguing pulsar behavior.
 
\acknowledgments 

M.B., A.P. and N.D.A. received support from the Italian Ministry of
University and Research (MIUR) under the national program {\it Cofin
2003}. I.H.S. holds an NSERC UFA and is supported by a Discovery
Grant. F.C. acknowledges support from NASA grant NNG05GA09G. D.R.L. is a
University Research Fellow funded by the Royal Society.

%\bibliographystyle{apj} 
%\bibliography{journals,modrefs,psrrefs,crossrefs}

\end{document}